\documentclass[preprint]{aastex}

\usepackage{epsfig}

\newcommand{\unit}[1]{\ensuremath{\,\mathrm{#1}}}
\newcommand{\degs}{\ensuremath{^\circ~}}

\slugcomment{}

\shortauthors{Oser et al.}
\shorttitle{Observations Of The Crab With STACEE}

\begin{document}


\title{HIGH ENERGY GAMMA-RAY OBSERVATIONS \\
	OF THE CRAB NEBULA AND PULSAR WITH THE \\
	SOLAR TOWER ATMOSPHERIC CHERENKOV EFFECT EXPERIMENT}


\author{S. Oser\altaffilmark{1,2}, D. Bhattacharya\altaffilmark{3},
L. M. Boone\altaffilmark{4}, M. C. Chantell\altaffilmark{1},
Z. Conner\altaffilmark{1,5}, C. E. Covault\altaffilmark{1},
M. Dragovan\altaffilmark{1,6}, P. Fortin\altaffilmark{8},
D. T. Gregorich\altaffilmark{7},
D. S. Hanna\altaffilmark{8}, R. Mukherjee\altaffilmark{9},
R. A. Ong\altaffilmark{1}, K. Ragan\altaffilmark{8},
R. A. Scalzo\altaffilmark{1}, 
D. R. Schuette\altaffilmark{1}, 
C. G. Th\'eoret\altaffilmark{8},
T. O. T\"umer\altaffilmark{3}, D. A. Williams\altaffilmark{4},
and J. A. Zweerink\altaffilmark{3}}


\altaffiltext{1}{Enrico Fermi Institute, University of Chicago,
Chicago, IL 60637, USA.} 
\altaffiltext{2}{present address: Department of Physics and Astronomy,
University of Pennsylvania, Philadelphia, PA 19104, USA.} 
\altaffiltext{3}{Institute of Geophysics and Planetary Physics,
University of California, Riverside, CA 92521, USA.}
\altaffiltext{4}{Santa Cruz Institute for Particle Physics, University
of California, Santa Cruz, CA 95064, USA.} 
\altaffiltext{5}{present address: Department of Physics/Women and
Power Leadership Programs, The George Washington University,
Washington DC 20007, USA.}
\altaffiltext{6}{present address: Jet Propulsion Lab, Pasadena, CA
91109, USA.}
\altaffiltext{7}{Department of Physics and Astronomy, California State
University, LA, Los Angeles, CA 90032, USA.} 
\altaffiltext{8}{Department of Physics, McGill University, Montreal,
QC H3A~2T8, Canada.}
\altaffiltext{9}{Department of Physics \& Astronomy, Barnard College
\& Columbia University, New York, NY 10027, USA.}


\begin{abstract}
The Solar Tower Atmospheric Cherenkov Effect Experiment (STACEE) is a
new ground-based atmospheric Cherenkov telescope for gamma-ray
astronomy.  STACEE uses the large mirror area of a solar heliostat
facility to achieve a low energy threshold.  A prototype experiment
which uses 32 heliostat mirrors with a total mirror area of $\sim
1200\unit{m^2}$ has been constructed.  This prototype, called
STACEE-32, was used to search for high energy gamma-ray emission from
the Crab Nebula and Pulsar.  Observations taken between November 1998
and February 1999 yield a strong statistical excess of gamma-like
events from the Crab, with a significance of $+6.75\sigma$ in 43 hours
of on-source observing time.  No evidence for pulsed emission from the
Crab Pulsar was found, and the upper limit on the pulsed fraction of
the observed excess was $< 5.5\%$ at the 90\% confidence level.  A
subset of the data was used to determine the integral flux of gamma
rays from the Crab.  We report an energy threshold of $E_{th} = 190
\pm 60\unit{GeV}$, and a measured integral flux of $I (E > E_{th}) =
(2.2 \pm 0.6 \pm 0.2)~\times~10^{-10}\unit{photons~cm^{-2}~s^{-1}}$.
The observed flux is in agreement with a continuation to lower
energies of the power law spectrum seen at TeV energies.
\end{abstract}




\section{INTRODUCTION}

The Crab Nebula is the most studied object in the gamma-ray sky.
Numerous ground-based experiments have detected emission from the Crab
at energies from $300\unit{GeV}$ through $50\unit{TeV}$.  In addition,
the EGRET detector on the Compton Gamma Ray Observatory has measured
the spectrum of unpulsed gamma rays from the Crab up to an energy of
$\sim 10\unit{GeV}$ \citep{nolan93}.  High energy gamma rays are
believed to result from inverse Compton scattering of relativistic
electrons on synchrotron photons, thermal dust photons, and cosmic
microwave photons.  Such inverse Compton models have been described by
\citet{dh92}, \citet{aa96}, and \citet{dh96}.  Under this scenario,
high energy electrons are accelerated at the pulsar wind termination
shock.  They then propagate through the nebula, producing both
synchrotron and inverse Compton emission.

In addition to unpulsed gamma rays, EGRET has also
seen pulsed gamma-ray emission from the Crab
\citep{nolan93,raman95,fierro98}, which is assumed to be produced by
the pulsar.  Pulsed emission has not been seen at higher energies
accessible to ground-based experiments \citep{lessard99}.  The
measurements require that pulsed emission cut off somewhere between
$10\unit{GeV}$ and $\sim 250\unit{GeV}$.  Polar Cap models of pulsar
emission \citep{dh82} predict sharp cutoffs near $10\unit{GeV}$,
whereas Outer Gap models \citep{chr, romani96} predict pulsed emission
as high as $\sim 50\unit{GeV}$.

No previous gamma-ray detector has had sensitivity to gamma rays at
energies between $\sim 10\unit{GeV}$ and $\sim 300\unit{GeV}$.
Nonetheless, there are strong motivations to observe the Crab in this
energy range, not only to test inverse Compton models at previously
unexplored energies, but also to search for pulsed gamma-ray emission
that could distinguish between pulsar emission models.  Satellite
experiments have been limited to energies below $\sim 10\unit{GeV}$ by
low statistics because of their small apertures and the rapidly falling
flux of gamma rays.  Ground-based experiments, which detect gamma rays
by the Cherenkov light produced in extensive air showers initiated by
the gamma ray, typically have energy thresholds of $\sim
300\unit{GeV}$.  The energy threshold is limited by the ability of the
instrument to distinguish dim flashes of Cherenkov light amidst the
fluctuations in night sky background light.  From signal-to-noise
considerations, the energy threshold of an atmospheric Cherenkov
telescope can be shown to behave like
\begin{equation}
E_{th} \propto 
\sqrt{ \frac{\Phi \Omega \tau}{\epsilon A}}.
\label{eq.threshold}
\end{equation}
Here, $\Phi$ is the flux of night sky background light, $\Omega$ is
the field of view of the instrument, $\tau$ is the length of the
electronic trigger coincidence window, $\epsilon$ is the efficiency
for detecting Cherenkov photons, and $A$ is the total mirror
collection area of the instrument \citep{weekes88}.  In particular,
the number of Cherenkov photons collected from a shower will scale
with the mirror area $A$, while fluctuations in the night sky
background light will increase with the square root of the light
collection, and hence with $\sqrt{A}$.  Thus, the energy threshold of
an atmospheric Cherenkov detector decreases approximately in
proportion to the inverse square root of the total mirror area.

The Solar Tower Atmospheric Cherenkov Effect Experiment (STACEE) is a
new atmospheric Cherenkov detector for gamma-ray astronomy which uses
the large mirror area of a solar power research facility to achieve a
low energy threshold.  STACEE uses an array of steerable mirrors,
called heliostats, to collect Cherenkov light produced in extensive
air showers.  Heliostats focus the light onto secondary mirrors on a
central tower, which image the light onto photomultiplier tube cameras
(see Figure~\ref{fig.concept}).  Previous tests demonstrate that
heliostat arrays can be used to detect Cherenkov light from extensive
air showers \citep{ong96, chantell98}.  Here we report on observations
of the Crab Nebula and Pulsar taken with a prototype version of the
STACEE instrument.  This prototype, called STACEE-32, used 32
heliostats and a preliminary electronic trigger system.  The full
instrument incorporates 64 heliostats, and is presently under
construction.  It will start regular observations in the fall of 2000.

\section{DETECTOR}
\label{sec.detector}

STACEE is located at the National Solar Thermal Test Facility (NSTTF),
at Sandia National Laboratories in Albuquerque, NM, USA (34.962\degs
N, 106.509\degs W, 1705\unit{m} above sea level).  The NSTTF is a
solar power research facility of the U.S. Department of Energy.  It
includes 212 heliostat mirrors, each $\sim 37\unit{m^2}$ in area,
arrayed across a $300\unit{m} \times 120\unit{m}$ area (see
Figure~\ref{fig.stacee32_heliostats}).  For solar power research,
heliostats track the Sun and concentrate sunlight onto a central
receiver tower.  At night, STACEE-32 used a subset of the heliostats
to track a gamma-ray source.  These heliostats reflected Cherenkov
light towards two $1.9\unit{m}$ diameter secondary mirrors near the
top of the tower.  The secondary mirrors concentrated light from the
heliostats onto phototube cameras.  Each photomultiplier tube (PMT)
received light from a single heliostat in the array.  The heliostats
were tilted inwards to track ``shower maximum,'' the point of maximum
air shower development (about $11\unit{km}$ above sea level for
vertical showers).  Details of the site, heliostats, secondary
mirrors, and cameras are given elsewhere \citep{chantell98,oser}.

Each camera contained 16 photomultiplier tubes and received light from
a separate portion of the heliostat field (see
Figure~\ref{fig.stacee32_heliostats}).  Solid acrylic light collectors
based on the Dielectric Total Internal Reflecting Concentrator design
\citep{dtirc} were placed in front of the PMTs.  The collectors
concentrated incident light onto $5\unit{cm}$ diameter photomultiplier
tubes and restricted the field of view of each PMT on the secondary
mirror.  The field of view of a heliostat was determined by the
diameter viewed by its PMT on the secondary mirror, divided by the
distance between the secondary mirror and the heliostat.  The fields
of view were adjusted so that each heliostat saw a solid angle of
approximately $1.2 \times 10^{-4}$\unit{sr} on the sky.

The photomultiplier tubes (Photonis XP2282B model) had bi-alkali
photocathodes with peak sensitivity near $400\unit{nm}$ and excellent
time response ($\sim 1\unit{ns}$ resolution).  The single
photoelectron rate from night sky noise on each tube was $\sim
1.5\unit{GHz}$, and the typical operating gain was $1.1 \times 10^5$.
Signals from the PMTs were AC-coupled and amplified ($\times 100$) and
then routed to electronics racks inside the tower.

In order to achieve a low energy threshold, the PMT signals must be
combined in a short time coincidence (see eq.~[\ref{eq.threshold}]).
The heliostats are at varying distances from the tower, and light from
an air shower strikes different heliostats at different times,
depending on the orientation of the shower.  Therefore, signals from
different heliostats must be delayed by varying amounts in order to
put them into time coincidence.  STACEE-32 used a two-level digital
trigger to form this coincidence.  The 32 heliostats were divided into
four compact subclusters of eight heliostats each, as indicated in
Figure~\ref{fig.stacee32_heliostats}.  PMT signals were discriminated,
and the discriminator outputs were individually delayed to put the
eight signals within a subcluster into time coincidence with one
another.  If five of eight tubes fired within a 12.5\unit{ns} window,
then that subcluster triggered.  The four subcluster triggers were
themselves delayed and put into time coincidence.  Three of four
subclusters firing within a 15\unit{ns} window resulted in an
experiment trigger, which initiated readout.  Thus, a total of at
least fifteen heliostats had to fire to trigger the detector.  The
discriminator threshold was chosen so that the accidental trigger rate
due to night sky background fluctuations was negligible ($<
0.002$\unit{Hz}).  The typical discriminator level was $\sim
170\unit{mV}$, which was approximately 5.5 times the mean single
photoelectron amplitude.  The mean discriminator counting rate in each
channel was $\sim 2\unit{MHz}$.

Multi-hit time-to-digital converters (Lecroy 3377 TDC) measured the
arrival times of pulses from each discriminator channel.
Charge-integrating analog-to-digital converters (Lecroy 2249SG ADC) on
24 channels measured the pulse charge over a 37\unit{ns} integration
gate.  The long delay times in the trigger formation ($\sim 1\unit{\mu
s}$) required that the ADC inputs be first routed through $\sim
1200\unit{ns}$ of high quality RG213 coaxial cable.  These cables
delayed the arrival of the pulses at the ADC until after the trigger
had formed.  Scaler counters recorded the PMT and subcluster trigger
rates, and a GPS clock recorded the Universal Time for each event
trigger.  The PMT currents were also read out for every event.  An
experiment trigger asserted a veto, which inhibited further triggers
until the event was read out and the veto was cleared by the data
acquisition (DAQ) system.  Counters measured the livetime of the
experiment (the fraction of time for which triggers were enabled).
The experiment was also triggered with an external pulser every two
seconds to collect a sample of non-Cherenkov events for calibration
purposes.  All readout and control was done via an interface to a
Silicon Graphics workstation.  Details of the electronics are given in
\cite{oser}.

Calibration systems for STACEE-32 included a nitrogen laser and an LED
flasher circuit.  Optical fibers carried pulses from the laser to
small wavelength shifter plates mounted at the center of each
secondary mirror.  The laser light pulsed all phototubes in each
camera simultaneously.  A filter wheel was used to attenuate the laser
light level, and the slewing responses of the PMTs (time versus pulse
amplitude) were so determined.  On an occasional basis, the LED
flasher was placed in front of each PMT to measure its gain {\em in
situ}.  Phototubes, delay modules, and other electronics systems were
also calibrated extensively in the lab before installation.  In
addition, the orientations and status of all heliostats were logged
during tracking, and any malfunctions were noted offline and the
affected data were identified.

\section{EVENT RECONSTRUCTION AND BACKGROUND REJECTION}
\label{sec.reconstruction}

Although the energy threshold of an atmospheric Cherenkov detector is
ultimately limited by night sky background light, the sensitivity of
the detector is limited by a different kind of background, namely,
cosmic ray air showers.  STACEE-32's high-multiplicity trigger
coincidence effectively eliminated all accidental triggers due to
fluctuations in night sky light.  The high multiplicity of the trigger
also rejected approximately 98\% of all Cherenkov showers initiated by
cosmic rays (see \S~\ref{sec.hadronic_rejection}).  The remaining
cosmic ray triggers form the background from which gamma-ray events
must be distinguished.

Traditional atmospheric Cherenkov telescopes use a single mirror with
a pixellated phototube camera.  Each element in the camera is mapped
to a different angular region on the sky, and thus one records an
image of the Cherenkov shower in the sky.  However, STACEE is not an
imaging detector.  Instead, the experiment is a lateral array which
samples the Cherenkov wavefront at many locations within the light
pool.  In this respect, event reconstruction must be handled quite
differently than for an imaging Cherenkov telescope.  STACEE records
the arrival time and photon density of the Cherenkov wavefront at each
heliostat for every event.  From these quantities the shape and
lateral density profile of the Cherenkov light pool are determined.
For STACEE-32, the charge resolution on the ADC measurements was
limited by the high night sky light levels and relatively long
integration gates.  The charge resolution had a typical value of 8
photoelectrons.  This resolution was not adequate for reconstructing
the lateral density profile for most events, and so the ADCs were used
only for slewing corrections and diagnostics.  The full STACEE
detector will use 1\unit{GHz} waveform digitizers, which will provide
greatly improved resolution and allow reconstruction of the lateral
density profile.  For STACEE-32, event reconstruction was based solely
upon timing information.

\subsection{Timing Reconstruction}
\label{sec.timing}

Multi-hit TDCs measured the arrival times of PMT pulses.  For each
channel, the signal propagation times through the electronics were
calibrated, and thus the expected arrival time of each Cherenkov pulse
could be calculated, relative to the trigger time.  We defined a time
window with 12\unit{ns} width centered around the expected hit
location.  A heliostat which had a TDC hit within its time window is
said to have an ``in-time hit.''  Some channels did not have in-time
hits, if those heliostats were not hit by the air shower, or if the
Cherenkov pulse failed to exceed the discriminator threshold.  For
every in-time hit, we reconstructed the arrival time of the Cherenkov
pulse at the heliostat by correcting for the transit time of light
between the heliostat and the PMT, and the calibrated transit time of
the PMT pulse through the electronics.  ADC measurements were used to
apply a slewing correction to those channels which had ADCs.  The
values of the slewing correction were determined from laser
calibrations.

From the shape and orientation of the reconstructed timing wavefront
at the heliostats, we determined the incident direction of the primary
which initiated the air shower.  At high energies ($E > 1\unit{TeV}$),
the Cherenkov-emitting core of the air shower extends through the
atmosphere like a line source, and the resulting Cherenkov wavefront
has a conical shape.  At lower energies, however, the shower does not
penetrate very far into the atmosphere, and most of the Cherenkov
light is produced near the location of shower maximum.  For this
situation, the air shower approximates a point source of Cherenkov
light, and the resulting wavefront has a more spherical shape.  A ray
drawn between the center of the light pool on the ground and the
center of this sphere will point back towards the gamma-ray source.

The timing wavefront of each triggering event was fit to a spherical
shape.  Only those heliostats which had an in-time hit and a
slewing-corrected time were used in the fit.  The timing resolution
for each channel was estimated from its discriminator rate on an
event-by-event basis.  Calibration runs taken with cosmic ray triggers
showed that the timing resolution for channel $i$ was related to its
rate $R_i$ by an empirical relation:
\begin{equation}
\sigma_{t,i} = \sigma_{0,i} \sqrt{1 + \left( \frac{R_i}{R_{0,i}} \right)^2}~.
\label{eq.timing_res}
\end{equation}
The constants $\sigma_{0,i}$ and $R_{0,i}$ were determined for each
channel from the calibration data.  Using the estimated timing
resolution, the corrected times were fit to a sphere by a least
squares fitting procedure.  The location of the sphere's center was
allowed to vary, but the radius of the sphere was constrained to a
distance corresponding to a slant depth of 271\unit{g~cm^{-2}} from
the top of the atmosphere.  This radius corresponds to the average
location of shower maximum for a 100\unit{GeV} gamma ray.  The fit was
done three times in an iterative manner, first excluding points lying
4 standard deviations from the original fit, and then excluding any
remaining points lying more than 4.5 standard deviations from the
second fit.

For STACEE-32, the angular resolution was dominated by uncertainties
in the determination of the center, or core, of the Cherenkov light
pool.  For gamma-ray air showers, the lateral density profile is quite
smooth, and it is difficult to determine its center.  For this
analysis, we assumed the core location for all events to be at the
geometric center of the array.  Monte Carlo simulations indicate that
STACEE-32 had an angular resolution of $\sim 0.25\degs$ for
200\unit{GeV} gamma rays.  The field of view of the instrument was
only $0.35\degs$ (half-angle), and therefore selecting events based on
the reconstructed direction would not result in any significant
improvement in the significance of a detection.  However, the mean
direction of air showers reconstructed by STACEE-32 was within
0.1\degs of the expected pointing direction, which verified the
absolute pointing of the instrument to that accuracy.

\subsection{Hadronic Rejection}
\label{sec.hadronic_rejection}

Traditional imaging atmospheric Cherenkov telescopes for gamma-ray
astronomy reject cosmic ray background events based on the orientation
and width of the shower image in the camera.  As a non-imaging
Cherenkov telescope, STACEE uses very different means for
statistically identifying and removing cosmic ray events.

Cosmic ray showers differ from gamma-ray showers in the total amount
of Cherenkov light they produce.  A gamma-ray shower produces, on
average, significantly more Cherenkov photons than a cosmic ray of the
same energy (see Figure~\ref{fig.cherenkov_species}).  This effect
happens because the nuclear cascade of a cosmic ray air shower
contains fewer Cherenkov-producing particles than the electromagnetic
cascade of a gamma-ray air shower, and because the energy threshold
for Cherenkov light production is higher for nucleons than for
electrons.  Furthermore, the difference in the Cherenkov yield between
gamma-ray showers and cosmic ray showers increases as the primary
energy decreases.  Gamma-ray air showers are brighter and trigger the
detector with much greater efficiency than cosmic ray showers.

In addition, there are large differences between the lateral density
profiles of the Cherenkov light in gamma-ray and cosmic ray events.
Air shower simulations show that gamma-ray showers produce smooth and
uniform light pools on the ground.  In contrast, cosmic ray showers
produce non-uniform light distributions.  In these showers, a large
proportion of the light is often concentrated in a small area on the
ground.  STACEE-32's trigger required 15 heliostats to be hit
simultaneously across the array, and it therefore imposed a uniformity
requirement.  The high multiplicity requirement of the trigger thus
selected the smooth, gamma-like events, and rejected most of the
irregular, hadron-like events.  Measured cosmic ray rates and Monte
Carlo simulations suggest a trigger rejection factor of $\sim 50$ for
cosmic rays, compared to gamma rays with the same energy spectrum.
Nonetheless, there were additional hadronic cuts that were applied
offline to further suppress the hadronic background.

Two hadronic cuts were developed for STACEE-32.  The first cut was a
trigger re-imposition cut.  After in-time hits were found for each
event (as described in \S~\ref{sec.timing}), the trigger multiplicity
requirement was re-imposed.  The in-time hit window (12\unit{ns} wide)
was smaller than the trigger coincidence width, and thus re-imposing
the trigger in software removed some events that passed the hardware
trigger.  Hadronic air showers have wider, more irregular timing
profiles than gamma-ray showers, and so re-imposing the trigger with a
narrower time window modestly increases the signal-to-noise for gamma
rays.  This cut also reduces the impact of any random PMT hits that
occur in coincidence with actual air showers.

The second event cut used the shape of the timing wavefront to
distinguish between gamma-ray showers and cosmic ray showers.  As
described above, gamma-ray showers have smooth, approximately
spherical, timing profiles.  Cosmic ray showers tend to have more
irregular, less spherical wavefronts (see Figure~\ref{fig.3d_shapes}).
The goodness-of-fit of a spherical shape to the wavefront was used to
distinguish between gamma rays and cosmic rays.  We selected events
based on the value of the chi-squared per degree of freedom
($\chi^2$/ndf) of the spherical fit used for angular reconstruction.
This selection should further increase the signal-to-noise of a
gamma-ray signal.

\section{OBSERVATIONS}

The STACEE-32 instrument observed the Crab extensively between 1998
November 15 and 1999 February 18.  Data were taken on clear and
moonless nights.  We required the Crab to be within 45\degs of zenith.
We took calibration data during times when Crab was low in the sky.
Weather conditions and the presence of clouds were monitored
regularly.

STACEE uses an ON-OFF observing strategy.  Off-source runs are used to
estimate the cosmic ray background level.  A signal for gamma-rays
shows up as an excess of ON events, compared to OFF events.  On-source
runs of 28 minutes length were taken with the heliostats tracking
the gamma-ray source.  Off-source runs were taken with the instrument
tracking a point in the sky displaced by one half hour in right
ascension from the gamma-ray source, but at the same declination.  For
each on-source run we took an off-source run of the same length.
Together these two runs form an ON-OFF pair.  The ON half of the pair
preceded or followed the OFF half by exactly one half hour, and so both
halves of the pair tracked the same trajectory in local coordinates on
the sky.  During these observations, STACEE-32 acquired a total of 141
ON-OFF pairs, corresponding to $\sim 65$ hours of on-source observing
time.

\subsection{Run Cuts}
\label{sec.run_cuts}
In order to do a valid background subtraction, the ON and OFF halves
of each pair must be closely matched in terms of detector properties
and weather conditions.  A series of run cuts were imposed to
remove pairs affected by changing conditions.

Run cuts were designed to remove data with poor or changing weather,
detector malfunctions, or high or fluctuating PMT rates.  Run pairs
affected by weather conditions were identified from observing
logs.  Given the relative proximity of the city of Albuquerque, clouds
reflect artificial lighting, and change the apparent sky brightness.
Observing logs identified six ON-OFF pairs taken under partially
cloudy conditions.  These were removed from the data set.  Another
pair was removed because atmospheric haze was noted.  Yet another pair
coincided with the arrival of a major high-pressure front, which
changed the atmospheric conditions on a short timescale.  Finally, on
a single night, frost developed on the heliostats, and four run pairs
from that night were discarded.  In all, twelve ON-OFF pairs were
removed because of weather conditions.

Major detector malfunctions affected a small number of runs.  Two
pairs had corrupted data because of failures in the DAQ system.  Runs
in which multiple heliostats malfunctioned in a single subcluster were
also discarded.  In these runs, the affected subcluster had a
relatively low efficiency for participating in the trigger because its
effective coincidence level was tightened from 5 of 8 tubes to 5 of 6
tubes.  Five pairs were removed on this basis.  Runs in which a single
heliostat malfunctioned, or in which two heliostats in different
subclusters malfunctioned, were left in the data set, although the
affected heliostats were removed from event reconstruction for both
halves of the ON-OFF pair, as described in \S~\ref{sec.pmt_cuts}
below.  Heliostat malfunctions were almost always repaired the
following day.

High PMT rates degrade data in several ways.  As
equation~[\ref{eq.timing_res}] shows, the timing resolution worsens
with increasing rate.  The rate of accidental coincidences, completely
negligible under normal operating conditions, increases rapidly with
rising PMT rates.  Finally, high rates increase deadtime in the
subcluster trigger delay electronics.  Runs with high rates were
identified from their subcluster trigger rates.  Typical runs have
subcluster rates of $\sim 1\unit{kHz}$.  All runs in which the mean
trigger rate for any subcluster exceeded 20\unit{kHz} were removed.
If the RMS variation of a subcluster's rate within the run exceeded
3\unit{kHz}, the run was also removed.  These stringent cuts ensure
that the accidental trigger rate remained $< 0.01\unit{Hz}$, and that
the deadtime in the trigger delay of each subcluster was $< 1\%$.
Twenty ON-OFF pairs were removed by these cuts.

Finally, the data were scanned for runs with abnormally low event
trigger rates.  A single anomalous run was identified with a trigger
rate about eight standard deviations below that of similar runs.  The
low rate is indicative of a detector malfunction, the cause of which
is under investigation.  The pair containing this anomalous run was
removed from the final data set.

Run cuts were applied in a blind fashion, before the significance of
any possible gamma-ray excess was determined, so as not to bias the
result.  After application of run cuts, 101 ON-OFF pairs remain in the
data set.  Table~\ref{tab.run_cuts} summarizes the various run cuts.

\subsection{PMT Cuts}
\label{sec.pmt_cuts}

It was not uncommon for a single heliostat or electronics channel to
malfunction.  Removing the entire run in this circumstance is rather
draconian, and unnecessarily reduces the data set.  Instead, bad
channels were identified on a run-by-run basis.  Offending channels
were then removed from the offline analysis in both the ON and OFF
halves of the run pair, and the data were analyzed as if the PMT were
simply turned off.  The most common problems were heliostat tracking
errors, which could be identified from the heliostat log files.
Removing the bad channels from both halves of a pair ensures that the
ON and OFF pairs are balanced in terms of detector response.

\section{RESULTS}

After run cuts, the final STACEE-32 data set consists of 101 ON-OFF
pairs, with a total on-source observing time of 155,335\unit{s} ($\sim
43$ hours).  Each run was processed separately, and event
reconstruction proceeded as described in \S~\ref{sec.reconstruction}.
For each event, the GPS event time, fitted shower direction, and
$\chi^2$/ndf of the spherical fit were found.  The livetime
fraction for each run was calculated, and the event totals for each run
were corrected for deadtime by dividing by the livetime fraction
(typically between 88\% and 92\%).

A gamma-ray signal shows up as an excess of events in the ON runs,
compared to the OFF runs.  As described in
\S~\ref{sec.hadronic_rejection}, two kinds of event cuts were used to
enhance any potential gamma-ray signal.  The first cut was to
re-impose the trigger multiplicity in software.  The second cut was
based on the $\chi^2$/ndf of a spherical fit to the wavefront's shape.
Monte Carlo simulations suggest that a cut of $\chi^2$/ndf$~< 1$ will
maximize the signal-to-noise.

\subsection{Unpulsed Emission}
\label{sec.dc_emission}

First we performed a search for unpulsed emission, presumed to
originate in the nebula of the Crab.  The number of events in the
on-source and off-source runs were tallied, with and without event
cuts.  Table~\ref{tab.total_dc} summarizes the results.  The raw data
show a strong excess of 4860 events, before event cuts, against a
background of $\sim 420,000$ events.  The Li-Ma statistical
significance \citep{li-ma} of this excess is $+5.27\sigma$.  Of the
101 ON-OFF pairs, 65 pairs show an excess in their raw rates, while
just 36 show a deficit.

Next, the trigger multiplicity was re-imposed.  The total number of
excess events after trigger re-imposition was 4551, with a background
of $\sim 350,000$, for a total significance of $+5.44\sigma$.  Seventy
of the 101 pairs showed an excess of events from the source.

Finally, a cut on $\chi^2$/ndf~$< 1$ was applied on top of the trigger
re-imposition cut.  The remaining number of excess events was 4062.
The background was reduced by a factor of approximately two, to $\sim
180,000$ events, and the total significance of the signal increased
markedly to $+6.75\sigma$.  Of the 101 pairs, 77 now show an excess of
events from the source region, while just 24 show a deficit.

The observed excess is statistically strong, being present at greater
than the five standard deviation level in the raw trigger rates alone.
Because STACEE is a new detector using a novel technique, there is the
question of whether this excess is actually due to a gamma-ray signal,
or whether it could be due to some unforeseen systematic effect that
is present in spite of our efforts to closely match ON and OFF halves
of each pair.  This question will ultimately be decided by
confirmation of the result by other experiments.  Detection of other
sources with this technique should provide further confirmation of the
method.  Nonetheless, there are consistency checks which can be
performed to strengthen the conclusion that STACEE-32 sees gamma rays
from the Crab.

The first and most powerful check is that the significance of the
excess increases as selection cuts are applied to the data.  Both the
trigger re-imposition cut, and especially the cut on the $\chi^2$/ndf,
increase the significance of the signal by amounts in agreement with
expectations from simulation.  In short, the observed excess behaves
just as a gamma-ray signal should.

Secondly, one can look at the distribution of pairwise significances.
The pairwise significance is defined as the observed excess or
deficit, in standard deviations, for each ON-OFF pair.  For a steady
source with a constant flux, the distribution of pairwise
significances should be normally distributed, with unit width and a
shifted mean.  As is seen in Figure~\ref{fig.pairwise}, the
distribution of pairwise significances for our data does have the
expected form.  This fact further supports the interpretation of the
excess as a gamma-ray signal.

In Table~\ref{tab.monthly}, the data have been broken down by month.
Since fewer runs were taken in January and February of 1999, these
months have been combined.  An excess with a significance of $\sim
4\sigma$ is seen in each era.  Also noteworthy is that the background
rate (OFF events per unit time) decreased by 20\% between the November
data and the December-February data.  We interpret this decrease as an
increase in the energy threshold of the experiment.  Closer
examination of the data shows that the rates were steady within the
month of November, and within December as well.  Whatever change
happened to the detector must therefore have occurred during the
intervening full Moon period during which STACEE did not operate.
Although the reason for this increase in threshold is not known with
certainty, we suspect that the angular pointing of the heliostats
drifted slightly out of alignment, decreasing the optical throughput.

\subsection{Pulsed Emission Search}

Having established the presence of a gamma-ray signal in our data, we
then carried out a search for pulsed emission in phase with the Crab
Pulsar's emission.  A GPS clock recorded the arrival time of every
event with an accuracy of $\sim 1\unit{\mu s}$.  The arrival time of
each event at Earth was corrected to barycentric dynamical time (TDB)
using the JPL Planetary and Lunar Ephemerides DEC-200 package
\citep{jpl}.  The corresponding radio phases were obtained from the
Jodrell Bank ephemeris \citep{jodrell}.  The phase values were
interpolated between monthly epochs by expanding the phase in a Taylor
series, and by requiring continuity of the phase, period, and time
derivatives of the period at the midway point between the two months.
The barycentering and ephemeris calculations were checked by applying
them to optical pulsar data recorded by the Whipple gamma-ray
collaboration.  For these data, we verified that the optical pulse was
extracted properly.

The arrival times of on-source events were then folded with the pulsar
phase to produce a phase histogram for the STACEE-32 data.
Figure~\ref{fig.phase} shows the phase histogram for on-source events
passing all event cuts.  No obviously significant structure is seen.
We have applied the H-test to test the uniformity of the phase
histogram \citep{h-test, h-test1}.  The test yields a value for the
$h$ parameter of 5.37, which has a chance probability of 0.12.  We
therefore conclude that STACEE-32's Crab phase histogram is consistent
with being uniform in phase.

Concluding that our data set contains no strong evidence of pulsed
emission, we then set an upper limit on the pulsed fraction of the
total observed excess.  Because the pulse profile of the pulsed
emission at STACEE's energies is unknown, we make an assumption about
the phase intervals in which the gamma-ray emission would occur.  We
assume that the pulsed emission at STACEE energies would occur in
phase with the pulsed emission seen by EGRET
\citep{nolan93,raman95,fierro98}.  We defined an ``on-pulse'' region
for the phase histogram, which includes an interval centered about the
main radio pulse at 0.94-0.04 of the phase histogram, and a second
interval about the intrapulse from 0.32-0.43.  Using the method of
\cite{helene}, we derive an upper limit on the pulsed fraction of $<
5.5\%$ of the observed excess, at the 90\% confidence level.
Table~\ref{tab.pulsed} contains the event totals for the on-pulse and
off-pulse regions.

\subsection{Energy and Integral Flux Result}

The energy threshold and integral flux for these data must be
determined from careful calibration and simulations of all parts of
the detector.  STACEE-32 used three simulation packages to model the
response of the instrument to gamma rays.  The well-established MOCCA
code was used to model extensive air showers and Cherenkov light
production \citep{mocca}.  A complete ray-tracing program followed the
paths of photons through the optics.  Finally, a detailed electronics
simulation which made use of a library of digitized PMT pulses was
used to model the performance of the PMTs and electronics.  All parts
of the simulations have been verified against calibration data.

For the purpose of determining an energy threshold and integral flux,
only data taken in November 1998 were used.  As explained in
\S~\ref{sec.dc_emission}, these data had the lowest energy threshold,
as measured by the trigger rates.  Also, the detector was carefully
aligned during this month.  During the subsequent months, it is
believed that small degradations in the alignment increased the energy
threshold, but we do not have enough information to track the effect
reliably.  Using only data from November 1998 therefore provides the
lowest possible energy threshold and minimizes possible systematics
associated with optical alignment.  The increased statistical
uncertainty from using a subset of the data as opposed to the entire
data set is still smaller than the systematic uncertainty on the flux.
Likewise, we have used the total rates before event cuts in
determining the energy threshold and integral flux, since the
efficiencies of these cuts depend on the simulation, and so would
introduce additional systematic uncertainties if included.

Monte Carlo simulations were used to determine the sensitivity of
STACEE-32 to gamma rays.  The sensitivity was calculated for gamma
rays of various energies coming from multiple positions along the
Crab's trajectory across the sky.  The results were expressed as an
effective area curve, which is the gamma-ray collection area of
STACEE-32 as a function of the gamma-ray energy.  The average effective
area curve for the November 1998 data was calculated by weighting each
position on the sky by its exposure in the data set.
Figure~\ref{fig.eff_area} shows the average effective area as a
function of energy.  The sensitivity of STACEE-32 starts at an energy
below 100\unit{GeV}, and quickly rises with energy.  The effective
area approaches a plateau value of $\sim 28,000\unit{m^2}$ above
1\unit{TeV}.

To determine the energy threshold and integral flux, we assume that
the Crab's energy spectrum follows a differential power law of the
form
\begin{equation}
\frac{dN}{dE} = C E^{-2.4}.
\end{equation}
The power law index of $-2.4$ is estimated from measured Crab spectra,
interpolated to STACEE-32's energy range \citep{hillas98}.  Convolving
this spectrum with the effective area curve yields the differential
trigger rate per unit energy.  The peak rate occurs at an energy of
190\unit{GeV}.  This is our estimate for the so-called ``spectral
threshold energy.''  The energy threshold changes by less than $\pm
10\unit{GeV}$ if the assumed spectral index is varied by $\pm 0.2$.

Estimates of the uncertainty in each element of the simulations
(e.g. uncertainties in measured mirror reflectivities, PMT gains,
etc.) yield a total systematic uncertainty on the energy of 32\% ($\pm
60\unit{GeV}$).  No single factor dominates the systematic
uncertainty.

For the November data, a total of 183,501 on-source events were seen
in the raw data (without event cuts), versus 181,349 off-source
events, in 56,056\unit{s} of observing time.  The excess rate is
therefore $0.038 \pm 0.011\unit{Hz}$.  By integrating the assumed
differential flux with the effective area curve, and equating this to
the observed rate, we determine the integral flux.  We find that the
integral flux of gamma rays from the Crab Nebula above our energy
threshold of $E_{th} = 190 \pm 60\unit{GeV}$ is
\begin{equation}
I (E > E_{th}) = (2.2 \pm 0.6 \pm 0.2)~
\times~10^{-10}\unit{photons~cm^{-2}~s^{-1}}.
\end{equation}
Here the first error is statistical, and the second error is the
systematic error on the flux itself, not including the effects of
uncertainty in the energy threshold.  The dominant systematic error is
the uncertainty in the energy threshold $E_{th}$, which does not
change the flux value itself, but which does change the energy at
which that flux is reported.

\section{DISCUSSION}

Figure~\ref{fig.int_flux} shows STACEE-32's integral flux value.  Also
shown for comparison are measurements from the CAT \citep{cat} and
Whipple \citep{hillas98} experiments.  Our measured flux is consistent
with an extension to lower energies of the power law spectrum seen by
Whipple.  Inverse Compton models generally predict a hardening of the
spectral slope as the energy decreases.  Given the systematic
uncertainties, our measurement is consistent with these expectations.

No evidence of pulsed emission from the Crab Pulsar is seen, and an
upper limit on the pulsed fraction is derived at $< 5.5\%$ (90\% CL)
of the observed signal.  At present, the energy threshold of this
limit, although lower than any previous limit, is not low enough to
differentiate between Polar Cap and Outer Gap models.  As
the energy threshold of STACEE is lowered further, however, future
limits should constrain the theoretical models.

This detection is the first for the STACEE instrument.  The CELESTE
collaboration, which is operating its own solar heliostat experiment
at the Th\'emis site in France, has reported a preliminary detection
of gamma rays from the Crab using a similar technique.  Their data
analysis is in progress \citep{celeste_crab}.  Although STACEE-32 was
a prototype instrument, it achieved an unprecedentedly low energy
threshold.  The full STACEE detector will be completed in the year
2000, and will feature twice as many heliostats, improved trigger
electronics, and 1\unit{GHz} sampling waveform digitizers.  Based upon
the STACEE-32 results, we expect that the full STACEE instrument will
obtain its design goal of an energy threshold of $\sim 50\unit{GeV}$.

In the future, solar heliostat experiments such as STACEE, CELESTE
\citep{celeste}, and Solar Two \citep{solar2} will provide complete
spectral coverage at energies between $\sim 50$ and $500\unit{GeV}$
for the Northern Hemisphere.  These sorts of low threshold experiments
will provide continuity between satellite measurements and
ground-based detectors at TeV energies, and will naturally complement
future satellite experiments such as GLAST.



\acknowledgments

We are grateful to the staff at the NSTTF for their excellent support.
We thank the Physics Division of Los Alamos National Laboratory for
loans of electronic equipment.  We also thank the SNO collaboration
for providing us with acrylic for light concentrators, and the machine
shop staffs at Chicago and McGill for their assistance.  The Whipple
collaboration kindly provided us with optical pulsar data for testing
our barycentering analysis.  Many thanks to Tom Brennan, Katie Burns,
Jaci Conrad, Jim Hinton, William Loh, Anthony Miceli, Gora Mohanty,
Alex Montgomery, Heather Ueunten, and Fran\c{c}ois Vincent.  This work
was supported in part by the National Science Foundation, the Natural
Sciences and Engineering Research Council, FCAR (Fonds pour la
Formation de Chercheurs et l'Aide \`a la Recherche), the Research
Corporation, and the California Space Institute.  CEC is a Cottrell
Scholar of Research Corporation.

\clearpage



\figcaption[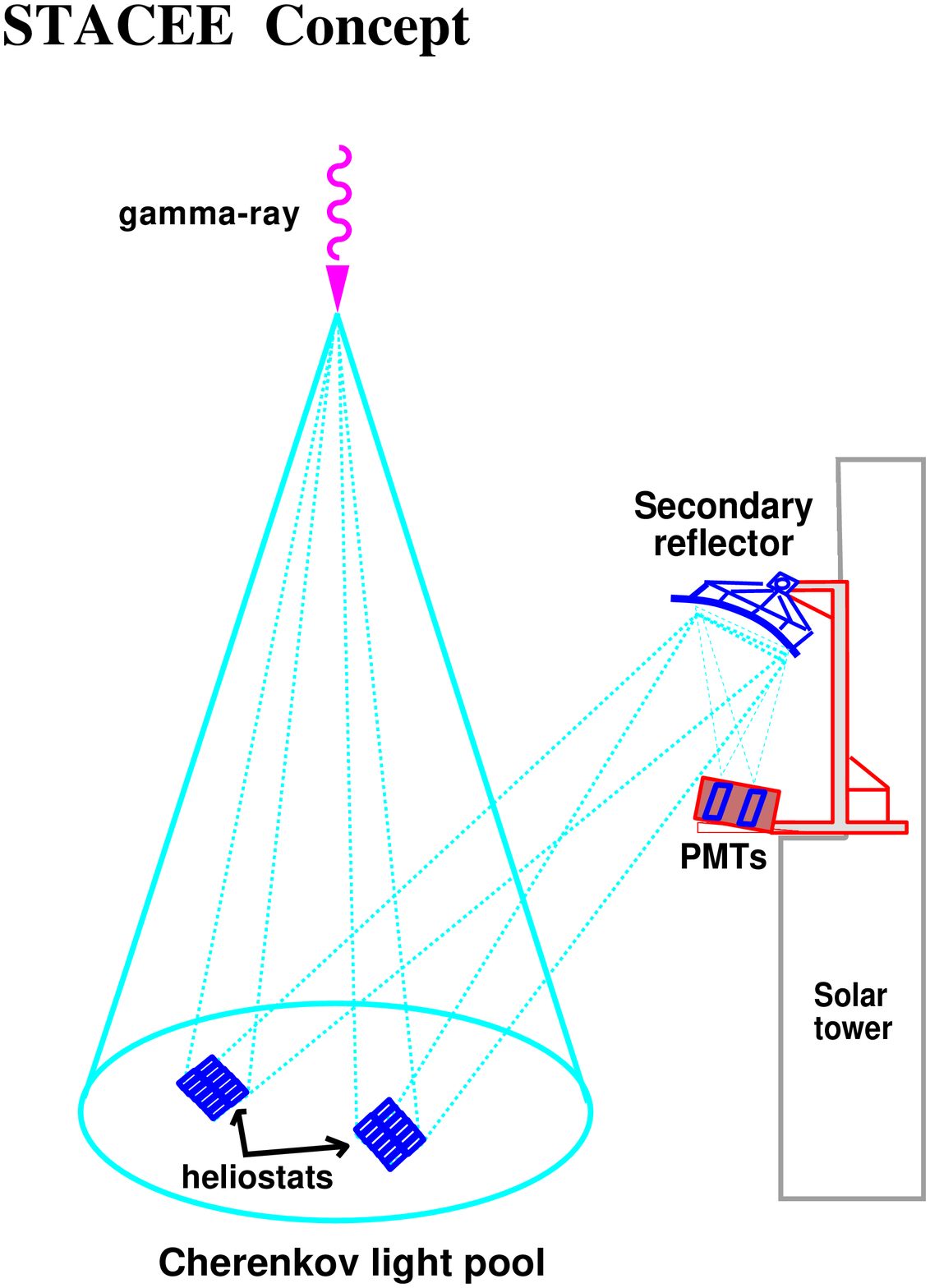]{STACEE Detector Concept Drawing.  An array of
heliostat mirrors collects Cherenkov light produced in a gamma-ray air
shower.  The heliostats reflect the light to secondary mirrors located
on a central tower.  The secondary mirrors image the light from each
heliostat onto an individual photomultiplier tube.  The STACEE-32
prototype used 32 heliostats and two secondary reflectors.  This
diagram is not to scale.
\label{fig.concept}}

\figcaption[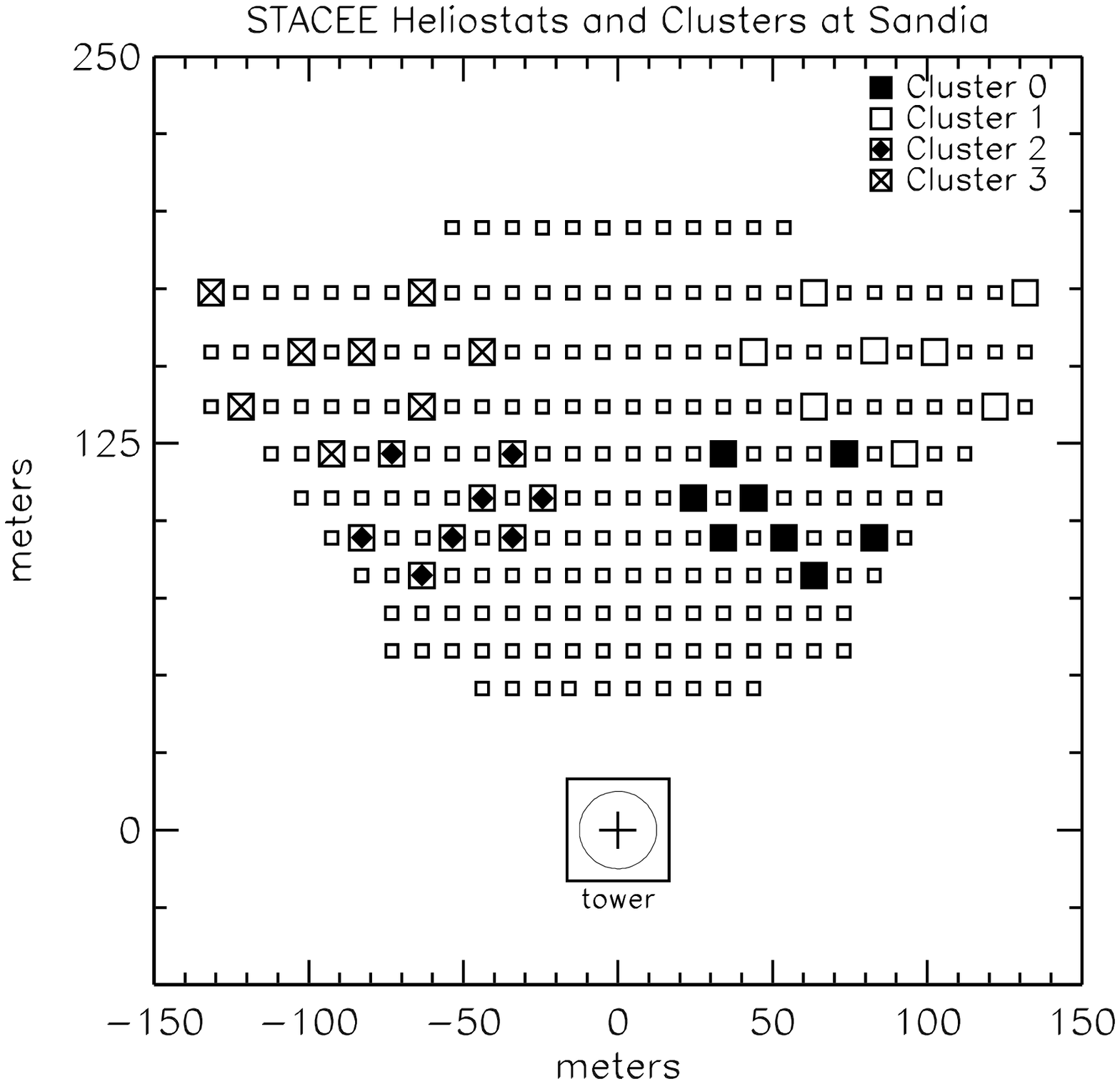]{Arrangement of Heliostats at the National Solar
Thermal Test Facility.  Two secondary mirrors (not shown) on the
central tower view sixteen heliostats each.  One secondary mirror
views the east half of the field, and one views the west.  The four
subclusters of eight heliostats each correspond to the four trigger
groups described in \S~\ref{sec.detector}.
\label{fig.stacee32_heliostats}}

\figcaption[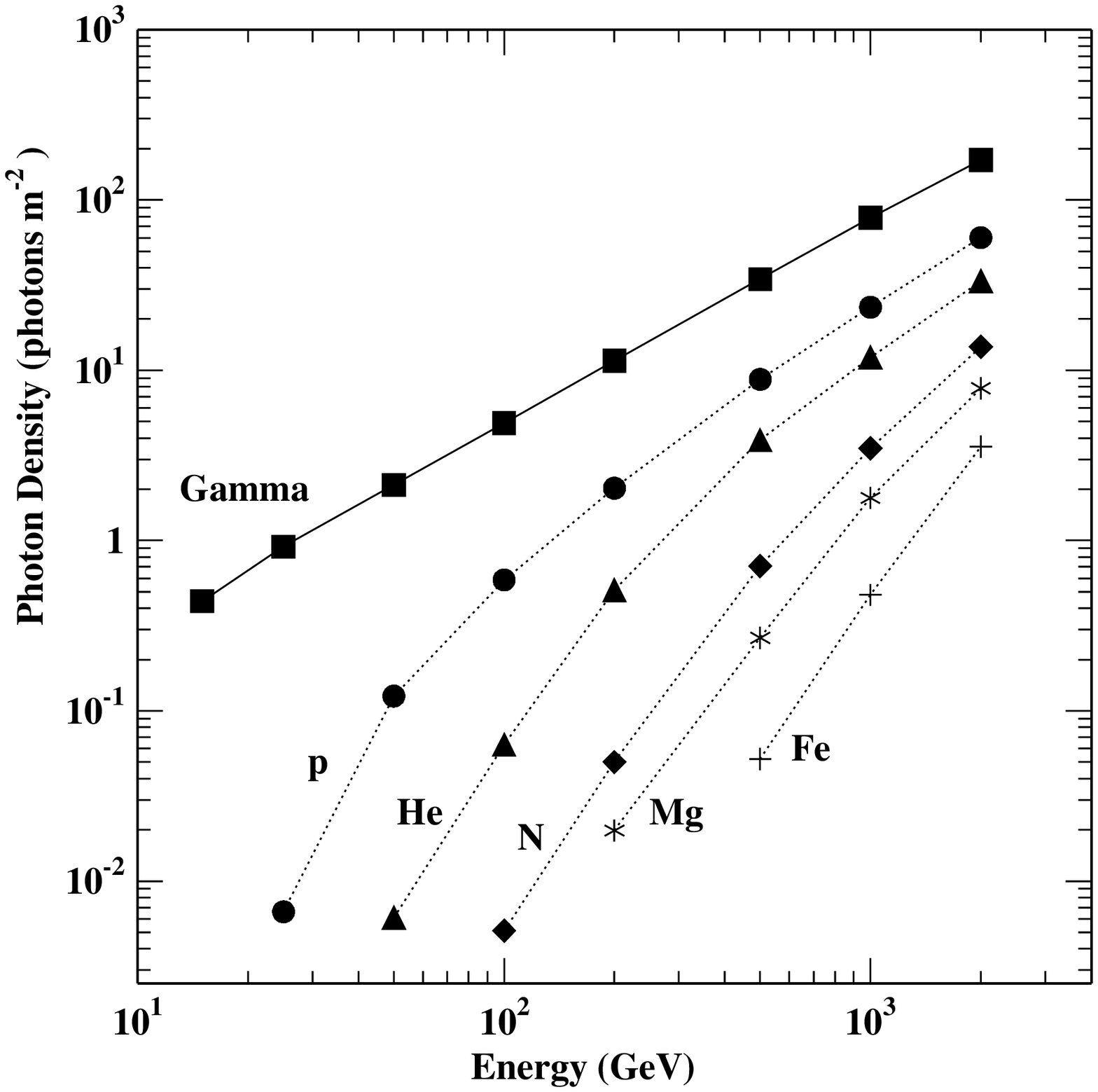]{Cherenkov Photon Yields versus Energy for
Different Species.  Plotted is the mean Cherenkov photon density
within 125\unit{m} of the shower core for vertically incident showers.
Only photons with wavelengths between 300-550\unit{nm} which land
within a 10\unit{ns} window around the peak arrival time are included.
\label{fig.cherenkov_species}}

\figcaption[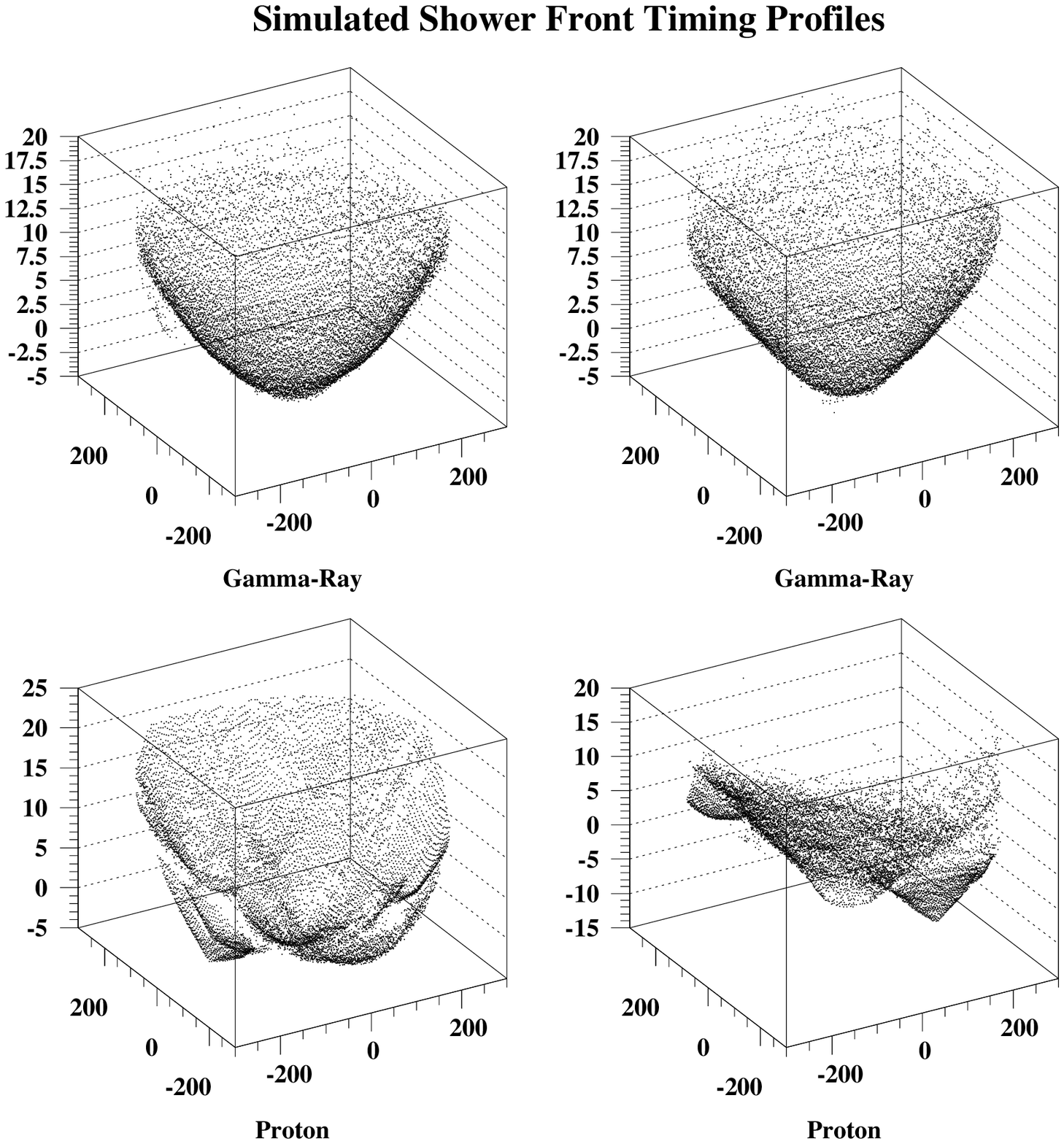]{Timing Wavefront Shapes for Simulated Gamma-Ray
and Proton Showers.  Plotted is the arrival time of the Cherenkov
photons on the ground, in nanoseconds, versus the location on the
ground, in meters.  Gamma-ray showers are approximately spherical near
the center of the shower (becoming more conical towards the edges),
while cosmic rays produce irregular, less spherical Cherenkov
wavefronts. \label{fig.3d_shapes}}

\figcaption[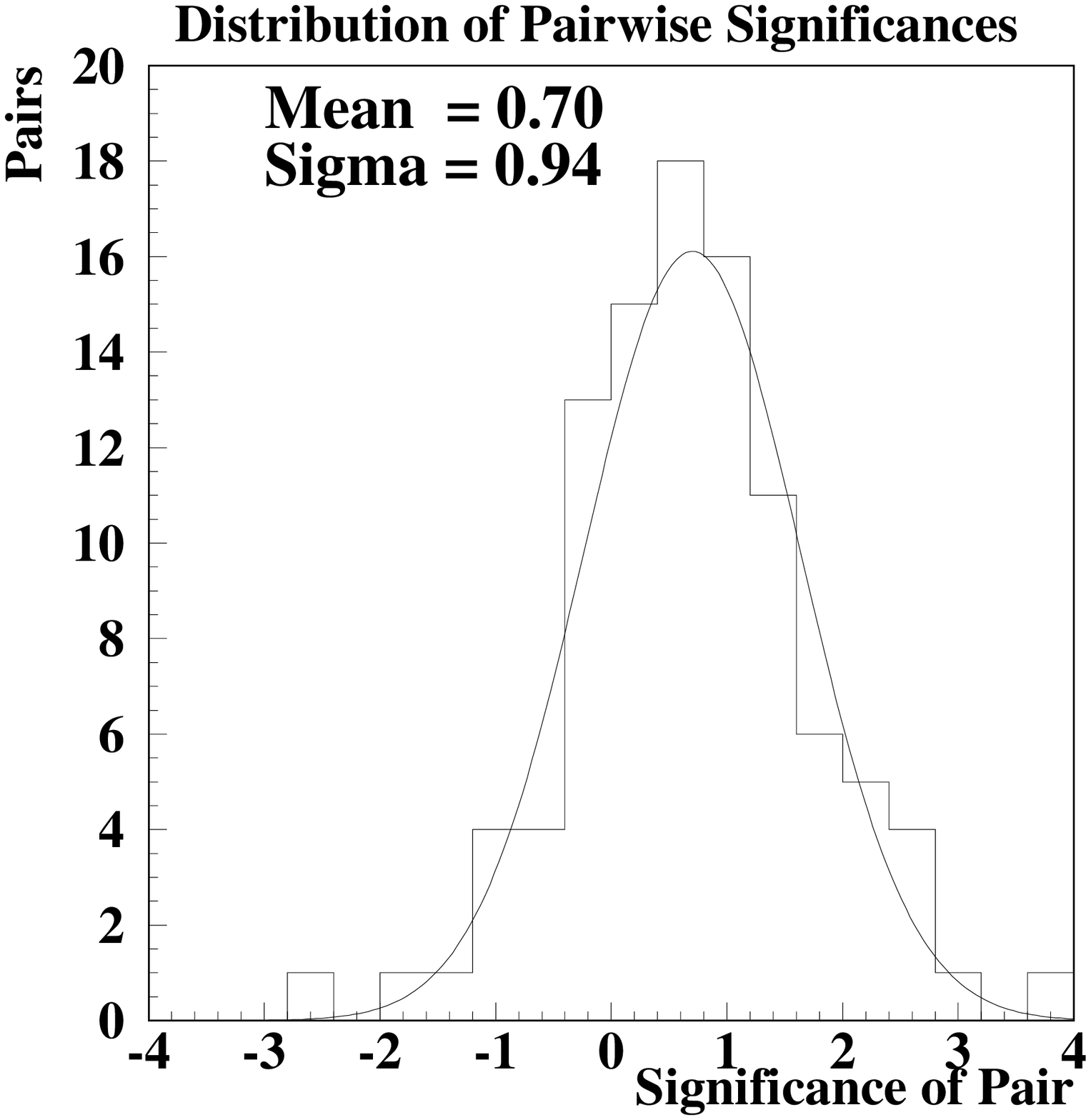]{Distribution of Pairwise Significances for the
STACEE-32 Crab Data.  The pairwise significance is defined as the
observed excess or deficit, in standard deviations, for each ON-OFF
pair.  For a steady gamma-ray signal, the distribution of pairwise
significances should be a Gaussian distribution with unit width.  The
significances are for event totals after all event cuts.
\label{fig.pairwise}}

\figcaption[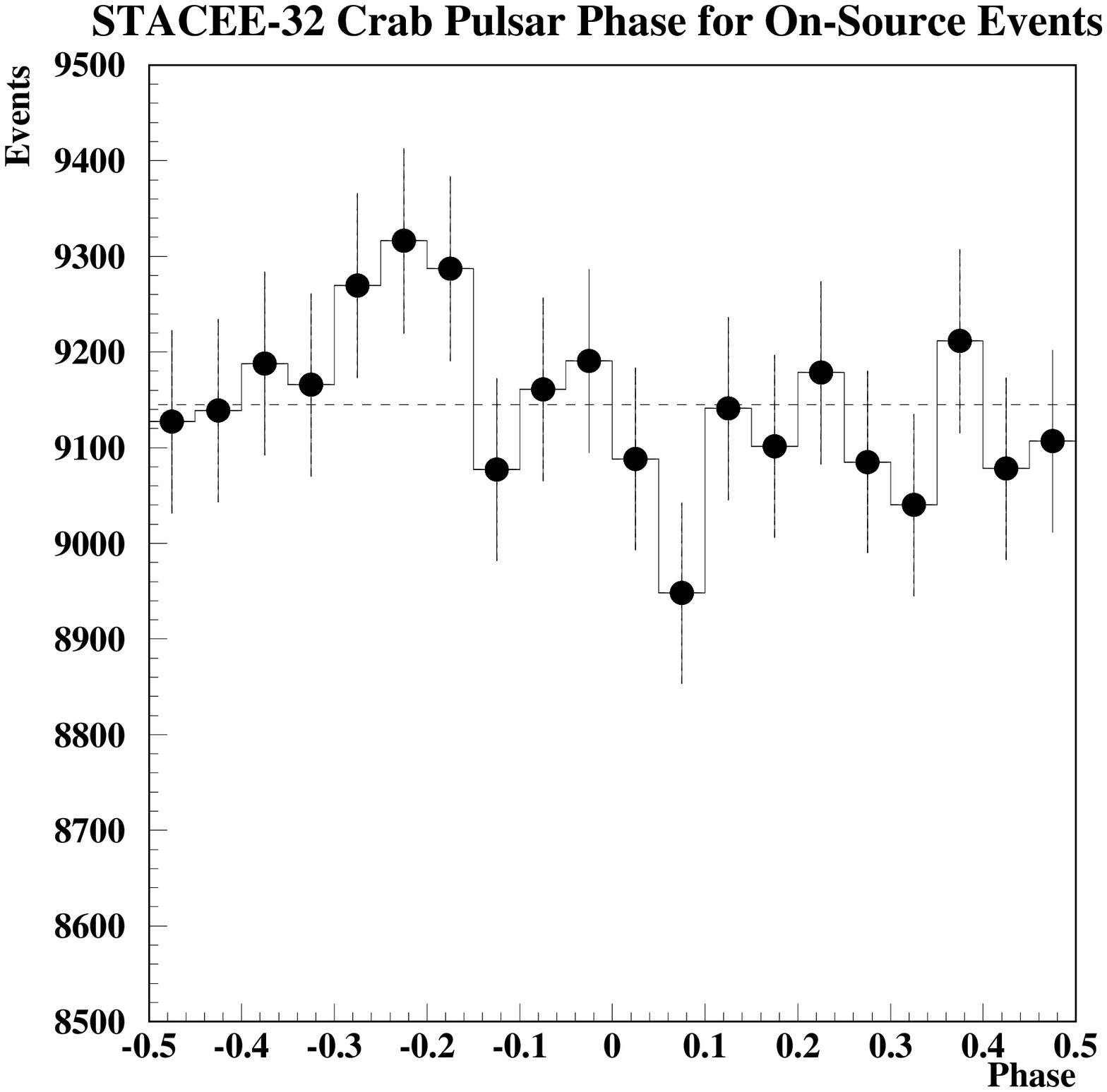]{ STACEE-32 Crab Pulsar Phase Histogram.  The
arrival times of on-source events are binned according to the radio
phase of the Crab Pulsar.  This plot includes all events passing the
trigger re-imposition and shower sphericity event cuts.  The main radio
pulse occurs at a phase of zero.  The horizontal dashed line is a fit
to a uniform phase distribution.
\label{fig.phase}}

\figcaption[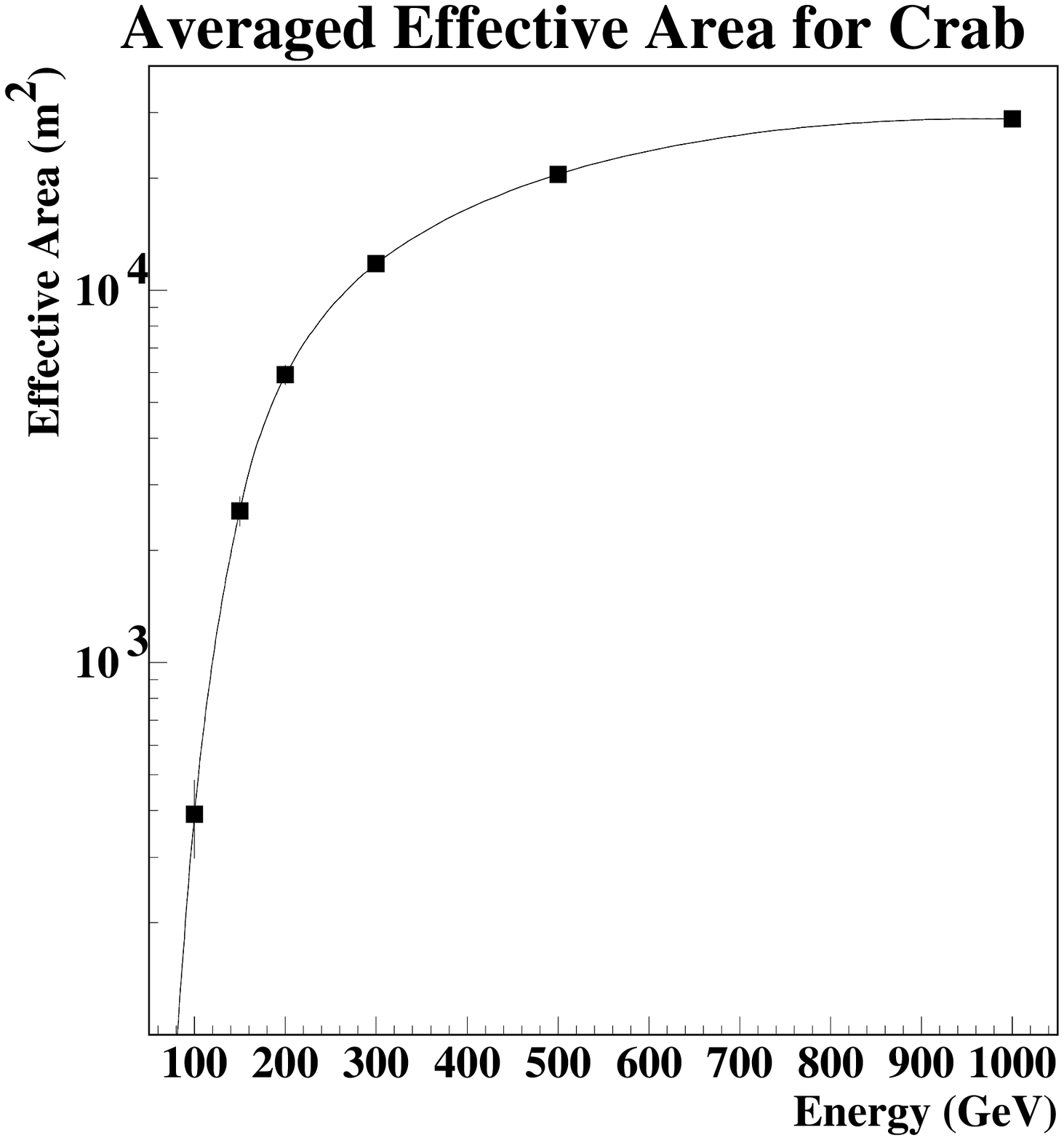]{ Average Effective Area Curve for
November 1998 Crab data set.  The error bars are statistical
only.\label{fig.eff_area}}

\figcaption[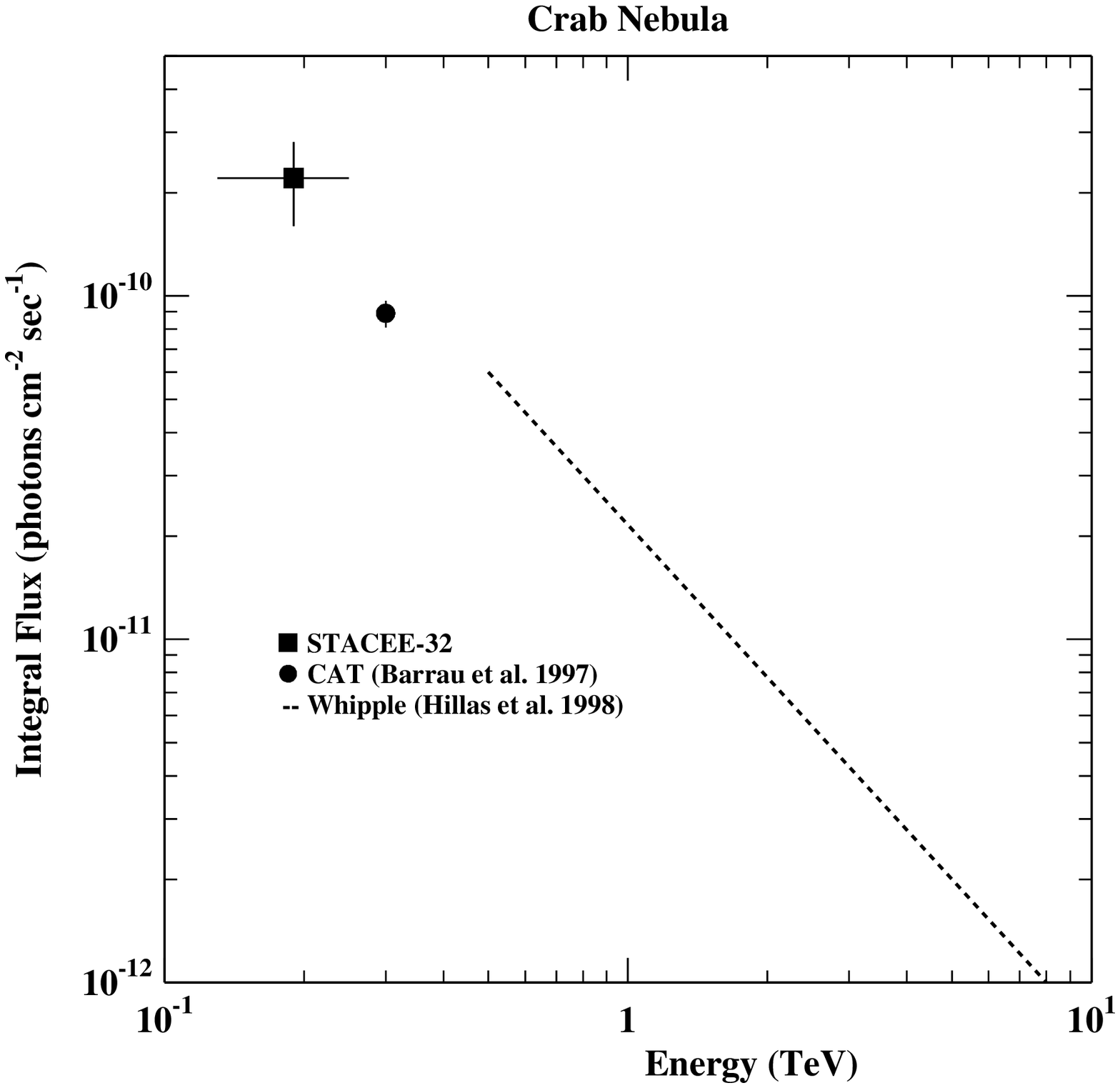]{STACEE-32 Integral Flux Result for Crab Nebula.
Also shown for comparison are the integral flux from the CAT
experiment and the measured spectrum from the Whipple Observatory.
The error bars on the STACEE-32 point include systematic error on the
energy threshold, and systematic and statistical errors on the flux.
The CAT data point includes only statistical error on the
flux. \label{fig.int_flux}}





\clearpage

\begin{table} 
\begin{center}
\caption[Summary of Run Cuts]
{Summary of Run Cuts.\label{tab.run_cuts}}
\begin{tabular}{lr}
\tableline
\tableline
Total pairs & 141 \\
\tableline
{\bf Runs Cut:} & \\
Clouds & 6 \\
Sky Clarity & 1 \\
Pressure Front & 1 \\
Frost & 4 \\
DAQ Malfunctions & 2 \\
Major Heliostat Malfunctions & 5 \\
High/Fluctuating Subcluster Rates & 20 \\
Abnormally Low Rate & 1\\
\tableline
Remaining Pairs & 101 \\
\tableline
\end{tabular}
\end{center}
\end{table}
\begin{table}
\begin{center}
\caption[ON$-$OFF Excesses for Crab Data Set]
{ON$-$OFF Excesses for Crab Data Set.  \label{tab.total_dc}}
\begin{tabular}{rccc}
\tableline
\tableline
           &    Raw   &    Trigger &    $\chi^2$/ndf \\
Quantity   &    Data  &    Re-imposed &    Cut \\
\tableline
   ON Events &     426975 &    352030 &    182915 \\
   OFF Events &    422115 &    347479 &    178853 \\
\tableline
   Excess (ON$-$OFF) &    4860 &    4551 &    4062 \\
   Significance &    +5.27$\sigma$ &    +5.44$\sigma$ &   
+6.75$\sigma$\\
   Pairs $+/-$ &    65/36 &    70/31 &    77/24 \\
\tableline
\end{tabular}
\tablecomments{ON-OFF event tallies are
shown for the raw data (no event cuts, but corrected for deadtime),
after a trigger re-imposition cut, and after an additional cut on the
sphericity of the shower wavefront.  ``Pairs $+/-$'' refers to how
many pairs showed an excess of events (+), versus a deficit
($-$).}
\end{center}
\end{table}
\clearpage
\begin{table}
\begin{center}
\caption[Monthly Crab Excesses] {Monthly Crab Excesses. \label{tab.monthly}}
\begin{tabular}{rcccc}
\tableline
\tableline
Quantity   &    Nov   &  Dec & Jan/Feb & Total \\
\tableline
On-source Time (s)   & 56056 & 51239 & 48040 & 155335 \\
ON Events            & 76235 & 55634 & 51046 & 182915 \\
OFF Events           & 74686 & 54342 & 49825 & 178853 \\
Significance     & $3.99\sigma$ & $3.90\sigma$ & $3.84\sigma$ & $6.75\sigma$\\Excess Rate (\unit{min^{-1}}) & $1.66 \pm 0.42$ & $1.51 \pm 0.39$ &
$1.52 \pm 0.40$ & $1.57 \pm 0.23$ \\
\tableline
\end{tabular}
\tablecomments{Results are
shown with all event cuts applied (trigger re-imposition and cut on
wavefront sphericity).  The January 1999 and February 1999 data have
been combined, since fewer runs were taken in these months.}
\end{center}
\end{table}
\begin{table}[b] 
\begin{center}
\caption[Pulsed Emission Search Results]
{Pulsed Emission Search Results. \label{tab.pulsed}}
\begin{tabular}{rc}
\tableline
\tableline
Number Of Events In Pulse Region      & 38173 \\
Number Of Events Outside Pulse Region & 144742 \\
Significance for Pulsed Region & $-1.37\sigma$ \\
Pulsed Fraction Of Emission (90\% C.L. Upper Limit) & $< 5.5\%$ \\
\tableline
\end{tabular}
\tablecomments{The phase search interval is: (0.94-0.04, 0.32-0.43).
The trigger re-imposition cut and a cut on the $\chi^2/$ndf for a
spherical fit have been applied.}
\end{center}
\end{table}


\clearpage
\nopagebreak
\begin{center}
\begin{figure}[t]
\figurenum{1}
\epsfig{file=f1.eps,width=5.5in}
\caption{}
\end{figure}
\end{center}

\begin{center}
\begin{figure}[p]
\figurenum{2}
\epsfig{file=f2.eps,width=5.5in}
\caption{}
\end{figure}
\end{center}

\begin{center}
\begin{figure}[p]
\figurenum{3}
\epsfig{file=f3.eps,width=5.5in}
\caption{}
\end{figure}
\end{center}

\begin{center}
\begin{figure}[p]
\figurenum{4}
\epsfig{file=f4.eps,width=5.5in}
\caption{}
\end{figure}
\end{center}

\begin{center}
\figurenum{5}
\begin{figure}[p]
\epsfig{file=f5.eps,width=5.5in}
\caption{}
\end{figure}
\end{center}

\begin{center}
\figurenum{6}
\begin{figure}[p]
\epsfig{file=f6.eps,width=5.5in}
\caption{}
\end{figure}
\end{center}

\begin{center}
\begin{figure}[p]
\figurenum{7}
\epsfig{file=f7.eps,width=5.5in}
\caption{}
\end{figure}
\end{center}

\begin{center}
\begin{figure}[p]
\figurenum{8}
\epsfig{file=f8.eps,width=5.5in}
\caption{}
\end{figure}
\end{center}


\end{document}